\documentclass[aps,prd,twocolumn,noeprint,showpacs,nofootinbib,superscriptaddress]{revtex4-2}

\usepackage{graphicx}
\usepackage{dcolumn}
\usepackage{bm}

\usepackage{graphicx}
\usepackage{color}
\usepackage{natbib}
\usepackage{epsfig}
\usepackage{float}
\usepackage{xcolor}
\usepackage{booktabs}
\usepackage{bibunits}

\usepackage{amsmath,amssymb,amsfonts}

\begin{document}

\preprint{APS/123-QED}

\title{Deep Residual Networks for Gravitational Wave Detection}

\author{Paraskevi Nousi}
\affiliation{Department of Informatics, Aristotle University of Thessaloniki, 54124 Thessaloniki, Greece}

\author{Alexandra E. Koloniari}
\affiliation{Department of Physics, Aristotle University of Thessaloniki, 54124 Thessaloniki, Greece}

\author{Nikolaos Passalis}
\affiliation{Department of Informatics, Aristotle University of Thessaloniki, 54124 Thessaloniki, Greece}

\author{Panagiotis Iosif}
\affiliation{Department of Physics, Aristotle University of Thessaloniki, 54124 Thessaloniki, Greece}

\author{Nikolaos Stergioulas}
\affiliation{Department of Physics, Aristotle University of Thessaloniki, 54124 Thessaloniki, Greece}

\author{Anastasios Tefas}
\affiliation{Department of Informatics, Aristotle University of Thessaloniki, 54124 Thessaloniki, Greece}

\date{\today}

\begin{abstract}

Traditionally, gravitational waves are detected with techniques such as matched filtering or
unmodeled searches based on wavelets. However, in the case of generic black hole binaries with non-aligned spins, if one wants to explore the whole parameter space, matched filtering can become impractical, which sets severe restrictions on the sensitivity and computational efficiency of gravitational-wave searches. Here, we use a novel combination of machine-learning algorithms and arrive at sensitive distances that surpass traditional techniques in a specific setting. Moreover, the computational cost is only a small fraction of the computational cost of matched filtering. The main ingredients are a 54-layer deep residual network (ResNet), a Deep Adaptive Input Normalization (DAIN), a dynamic dataset
augmentation, and curriculum learning, based on an empirical relation for the signal-to-noise
ratio. We compare the algorithm’s sensitivity with two traditional algorithms
on a dataset consisting of a large number of injected waveforms of non-aligned binary black hole mergers in real LIGO  O3a noise samples.
Our machine-learning algorithm can be used in upcoming rapid online searches of gravitational-wave events in a sizeable portion of the astrophysically interesting parameter space. We make our code, {\it AResGW}, and detailed results publicly available at \texttt{\url{https://github.com/vivinousi/gw-detection-deep-learning}}.

\end{abstract}

\maketitle


\section{Introduction}
\label{sec:intro}

In the first three observing runs (O1-O3) of the LIGO-Virgo Collaboration \cite{LIGO_ref, Virgo_ref} (recently joined by Kagra \cite{KAGRA_ref}) a number of ${\cal O} (90)$ confident gravitational wave (GW) detections were found in the data, including mostly binary black holes (BBH), but also a few binary neutron stars (BNS) and neutron star black hole systems (NSBH) \cite{GWTC1, GWTC2, GWTC3}. The anticipated significant increase in the number of detections during the fourth observing run (O4), which is scheduled to start in spring of 2023 and even more so during O5 and the observing runs of the planned 3rd-generation detectors (e.g. Cosmic Explorer \cite{2019BAAS...51g..35R} and Einstein Telescope \cite{2020JCAP...03..050M}) will make the application of traditional matched-filtering techniques increasingly costly or impractical \cite{2021arXiv211106987C}. This is both for reasons of computational efficiency, as the goal is to obtain near real-time detection triggers, as well as for reasons of accuracy, since the confident detection of near-threshold systems with random spin directions requires a much larger parameter space than the aligned-spin case. The situation will become even more challenging, if template banks with departures from general relativity (GR) will be included. Unmodeled search algorithms, on the other hand, naturally have a limited sensitivity, depending on the particular GW source.

An attractive solution to the above problem that has been investigated in the last few years is the implementation of machine-learning (ML) methods, such as convolutional neural networks (CNN) or auto-encoders,  see e.g. schafer2022training
\cite{PhysRevLett.120.141103,PhysRevD.97.044039,2019PhRvD.100f3015G,CORIZZO2020113378,PhysRevD.102.063015,2020PhRvD.101j4003W,2020PhLB..80335330K,2020arXiv200914611S,2021PhRvD.103f3034L,2021NatAs...5.1062H,2021MNRAS.500.5408M,2021PhLB..81236029W,2021PhRvD.104f4051J,10.3389/frai.2022.828672,2022arXiv220208671C,PhysRevD.105.043003,2022arXiv220606004B,schafer2022training,PhysRevD.106.042002,2022arXiv220704749A,2022arXiv220612673V}  and \cite{cuoco2020review} for a review, but it has been difficult to evaluate  the effectiveness of such efforts in a realistic setting. Recently, the first Machine-Learning Gravitational-Wave Mock Data Challenge (MLGWSC-1) was completed \cite{challenge1}, defining an objective framework for testing the sensitivity and efficiency of ML algorithms on modeled injections in both Gaussian and real O3a detector noise, in comparison to traditional algorithms. Here, we present the leading ML algorithm in the case of injections of BBH template waveforms in real O3a noise and show that with further improvements it surpasses, for the first time, the results obtained with standard configurations of traditional algorithms in this specific setting. This is achieved for a component mass range between $7-50 M_\odot$ (which corresponds to $70\%$ of the announced events in the cumulative GWTC catalog \cite{GWTC3}) and a relatively low false-alarm rate (FAR) as small as one per month. We thus demonstrate that our ML algorithm is sufficiently mature to be implemented in GW search pipelines.

\begin{figure*}[t!]
    \centering
    \includegraphics[width=\textwidth]{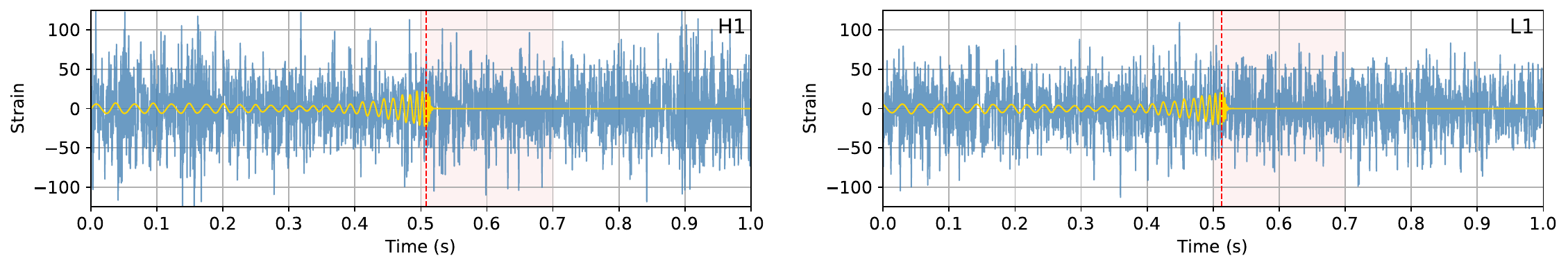}
    \caption{A representative 2-channel data segment of the training set containing an  injection in real O3a noise from the Hanford (H1) detector (left panel) and the Livingston (L1) detector (right panel). The {\it whitened strain} of a 1 s segment around the time of coalescence is shown. The coalescence times in the detector frames are within the 0.5s - 0.7s range (shown as a shaded area). The injected waveform is shown scaled to match the difference between the whitened foreground and background segments. In this example, the component masses are $m_1=27.74M_\odot$, $m_2=11.50M_\odot$ and
    the luminosity distance is $d=1497$Mpc. The {\it non-aligned spins} have magnitudes of 0.624 and 0.008, respectively.}
    \label{fig:injections}
\end{figure*}

A key indicator for the effectiveness of a GW search algorithm is the sensitive distance it can achieve at a given FAR \cite{PhysRevD.102.063015}. Our ML algorithm is a novel combination of several key ingredients, each of which boosts the sensitive distance to higher values. The base algorithm is a 54-layer one-dimensional (1D) deep residual network (ResNet) \cite{he2016deep}, which allows for training of a much deeper network, in comparison to the simpler CNN. The second important ingredient that allowed the algorithm to work well at small FAR is the addition of a
Deep Adaptive Input Normalization (DAIN)  \cite{dain}, to deal with the non-stationary nature of real O3a noise. Next, we obtained improvements by dynamically augmenting the dataset during training. The execution speed was significantly increased with the implementation of a framework-specific, module-based whitening layer, which computes the power spectral density (PSD) over a period of a few seconds in batched tensor format.
Finally, we saw a large boost in sensitive distance employing curriculum learning, in which the network was learning the injected waveforms with highest signal-to-noise ratio (SNR) first. 

The network was implemented in PyTorch \cite{PyTorch_ref} and its training (including validation) on 12 days of training data was achieved within 31 hours on an A6000 GPU with tensor cores (for 14 epochs). The runtime for the evaluation of one month of test data on the same hardware was less than 2 hours.

\section{Training and test datasets}
\label{sec:datasets}

The training dataset had a duration of 12 days and comprised real noise from the O3a LIGO run and injections of non-aligned BBH waveforms (corresponding to the assumptions of dataset 4 in \cite{challenge1}). The noise was sampled from parts of O3a that are  available from the Gravitational Wave Open Science Center (GWOSC). Only segments with a minimum duration of 2h, where both LIGO detectors registered good quality data, excluding 10s around detections listed in GWTC-2 were included (see \cite{challenge1} fore more details). Applying these criteria, leaves a dataset with noise for each of the two aLIGO detectors, Hanford (H1) and Livingston (L1), with a total duration of 11 weeks, with a sampling rate of 2048 Hz.

 The injected BBH waveforms were generated using the waveform model IMRPhenomXPHM \cite{IMRPhenomXPHM_ref}, in which a lower-frequency cutoff of 20 Hz was applied. The masses $m_1, m_2$ of individual  components were in the range  $7 M_\odot - 50 M_\odot$ (corresponding to a maximum signal duration of 20s). The signals were uniformly distributed in coalescence phase, polarization, inclination, declination and right ascension (see \cite{challenge1} for details). The waveforms were not injected uniform in volume, but the {\it chirp distance} $d_c$ was sampled (instead of the luminosity distance $d$). This choice increases the number of low-mass systems which can be detected. The chirp distance is defined as \begin{equation}
d_{c}=d\left(\frac{\mathcal{M}_{c, 0}}{\mathcal{M}_{c}}\right)^{5 / 6},
\end{equation}
where $
\mathcal{M}_{c} = \left(m_{1} m_{2}\right)^{3 / 5} /\left(m_{1}+m_{2}\right)^{1 / 5}$ is the chirp mass  and  $
\mathcal{M}_{c, 0}=1.4 / 2^{1 / 5} \mathrm{M}_{\odot}
$ is a fiducial value. The spins of the individual components had an isotropically distributed orientation  with a magnitude between 0 and 0.99 (hence, precession effects are present). All higher-order modes up to $(4,-4)$ available in  IMRPhenomXPHM are included. A taper window was applied to the start of each waveform. Figure \ref{fig:injections} illustrates a representative data segment of the training set.

The training dataset comprised the first 12 days of the 11-week dataset, resulting in 740k noise segments (for each of the two detectors) of duration 1.25s, 
which constituted the background noise. We then generated a set of 38k waveforms and each one was injected randomly into about 19 different background segments, resulting in 740k foreground segments that contain injections, leading to a balanced training set. The time of coalescence is chosen to randomly fall within  the 0.625s and 0.825s mark in each segment. This means that, regardless of the actual duration of the injected signal, only the last cycles corresponding to less than 1s duration are kept for training for any choice of parameters.

Similarly, we created a validation set, based on weeks 4 to 7 of the 11-week dataset, with a total duration of one month, comprising 1850k background noise samples and 1850k foreground samples containing injections (generated from 96k different injections, each injected randomly into about 19 different segments). 
The injections in the validation set were generated with a different random seed in comparison to the training set.  

The main test dataset comprises noise from weeks 8-11 of the 11-week dataset, with a duration of one month\footnote{A smaller test set of one day duration was generated to facilitate a faster comparison of different augmentation and training strategies.} and injections with merger times separated by a random time between 24 s to 30 s (which translates to about 96k injections over the whole duration). The code for creating the different datasets was provided by the organizers of MLGWSC-1 and is publicly available. For the main test dataset we used the same random seed of 2514409456 and offset of 0 (first 4 weeks) as in \cite{challenge1}. We created additional test datasets with the same total duration, but different random seeds and offset, in order to assess the variance of our main results.

\section{Computational methods}
\label{sec:methods}

\subsection{Data Preprocessing and Normalization}

Following approaches from traditional matched filtering as applied in GW detections, we first {\it whiten} the training data \cite{usman2016pycbc,PhysRevLett.120.141103}. As is common in deep learning (DL) methods, we then use an input normalization layer \cite{ioffe2015batch,dain},  before feeding the data to the neural network for classification.

{\it Whitening.}
The data segments were whitened using the Welch method \cite{villwock2008application} for computing the PSD and an inverse-spectrum truncation method \cite{allen2012findchirp} for smoothing. Both methods were implemented in PyTorch \cite{PyTorch_ref}, 
to allow for batch processing of multiple input segments in a single forward pass, which gave a significant speed up with respect to a CPU implementation.  First, we crop 4.25s segments from the original timeseries, for both the training and validation sets, with a stride of 1.4s. An injection is done in a random quarter segment (of 1.25s duration), which is whitened using the 4.25s noise segment surrounding it.  After whitening, the first and last 0.125s (0.25s 
in total) are removed from each sample, {\it leaving a 1 s sample as input for the neural network}. 


\begin{table}[]
    \centering
        \caption{Feature extraction backbone of our ResNet54-like architecture. The first column lists the number of residual blocks, totalling 27, each consisting of {\it two} convolutional layers. The second column lists the number of filters in each corresponding block. The third column indicates those blocks that are 2-strided (see text for details)  and the last column displays the dimensionality $D$ of the input tensor for each block.}
    \label{tab:resnet52_feature_extraction}
    \begin{tabular}{cccr}
    \toprule
        \hspace{0.2cm} residual blocks \hspace{0.2cm} & \hspace{0.2cm}  filters \hspace{0.2cm} & \hspace{0.2cm} strided \hspace{0.2cm} & input $D$ \hspace{0.2cm}\\ \midrule
        4 & 8 & & 2$\times$2048\\
        1 & 16 & \checkmark & 8$\times$2048 \\
        2 & 16 & & 16$\times$1024 \\
        1 & 32 & \checkmark & 16$\times$1024 \\
        2 & 32 & & 32$\times$512 \\
        1 & 64 & \checkmark & 32$\times$512 \\ 
        2 & 64 & & 64$\times$256 \\
        1 & 64 & \checkmark & 64$\times$256 \\ 
        2 & 64 & & 64$\times$128 \\
        1 & 64 & \checkmark & 64$\times$128 \\ 
        2 & 64 & & 64$\times$64 \\
        5 & 32 & & 64$\times$64\\
        3 & 16 & & 32$\times$64\\ 
        \bottomrule
    \end{tabular}

\end{table}

{\it Adaptive Normalization.}
As the real noise of the H1 and L1 detectors is non-stationary and affected by a multitude of factors, we opted for an {\it adaptive input normalization method}, specifically Deep Adaptive Input Normalization (DAIN) \cite{dain}, which is applied before feeding the data to the network. DAIN has been successfully used in tasks that involve non-stationary timeseries, such as financial timeseries forecasting \cite{passalis2021forecasting}.

\begin{figure}
    \centering
    \includegraphics[width=0.99\linewidth]{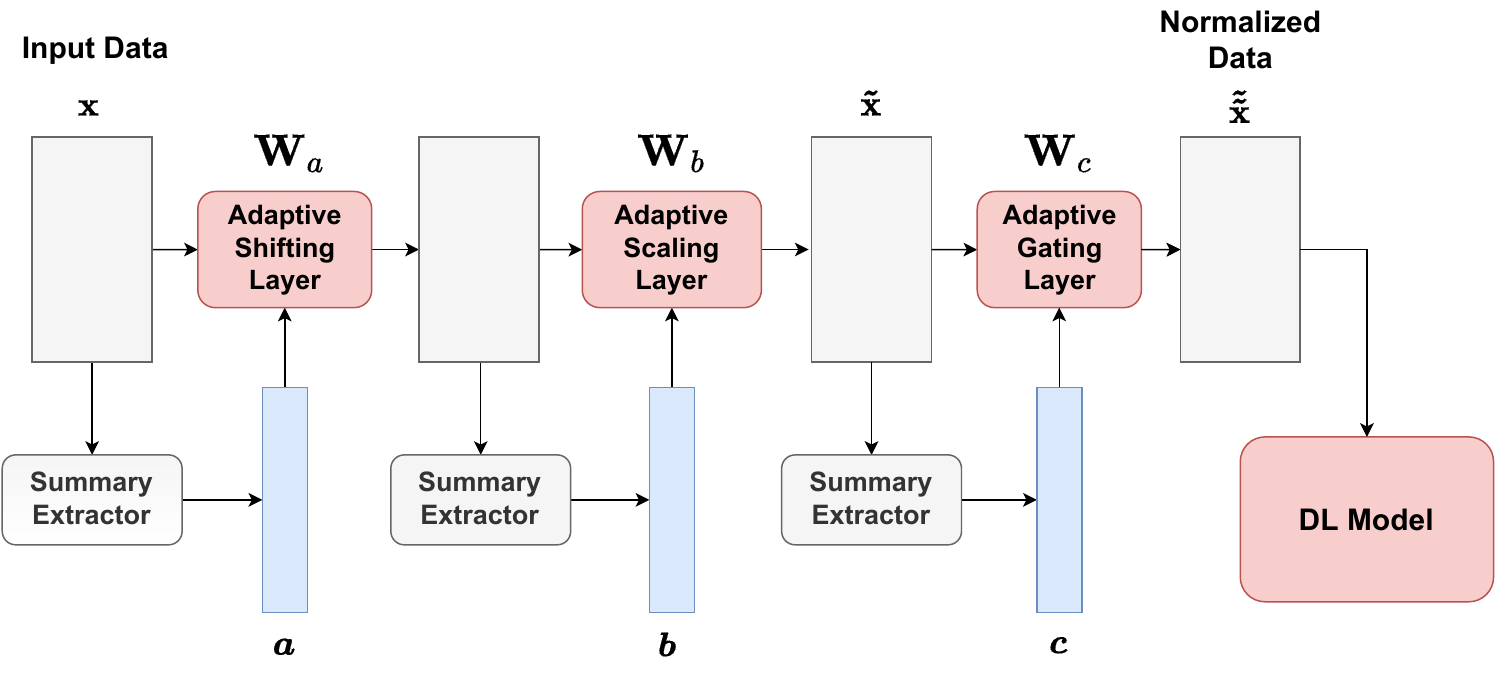}
    \caption{Deep Adaptive Normalization (DAIN) is applied before feeding the data to the employed DL model. DAIN involves three adaptive normalization steps, i.e., adaptive shifting, adaptive scaling and adaptive gating, increasing the ability of DL models to handle non-stationary data.  }
    \label{fig:dain}
\end{figure}

\begin{figure*}[t!]
    \centering
    \includegraphics[width=\textwidth]{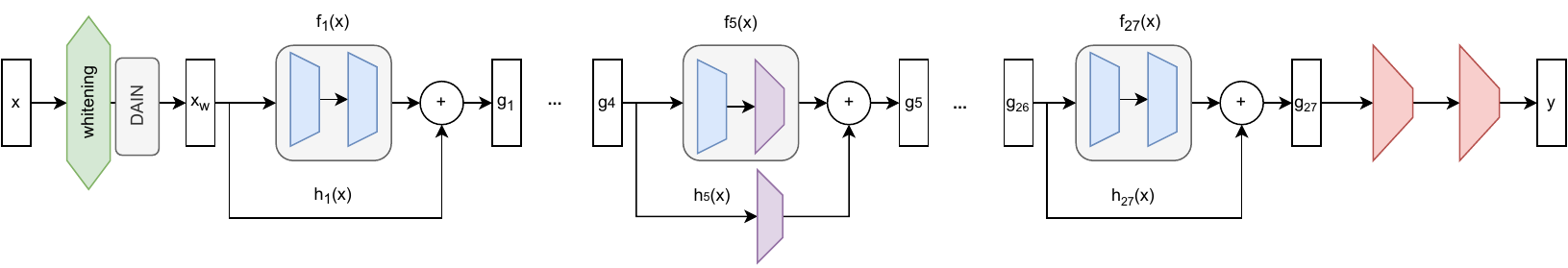}
    \caption{Description of our residual network architecture. The input $\mathbf{x}$ is $2\times 2048$-dimensional. There are 27 blocks comprising two convolutional layers. In five  of these blocks, the dimensionality is halved (stride 2, shown in purple) and an additional layer is used in the residual connection. Finally, there are two individual convolutional layers which gradually decrease the number of channels  down to two, corresponding to the binary classification targets (noise plus injected waveform vs. noise only) in the output.}
    \label{fig:resnet52}
\end{figure*}

The goal of this normalization layer is to \emph{learn} how the input timeseries should be normalized. DAIN is used before feeding the input to the subsequent layers and {\it it is trained by back-propagating the network's gradients} to its parameters, as shown in Fig.~\ref{fig:dain}. Furthermore, DAIN differs from other normalization schemes, since it can dynamically adjust the applied normalization scheme based on the input \emph{at inference time}, allowing for handling non-stationary data. 	More specifically, DAIN aims to {learn} how the measurements $\mathbf{x} \in \mathbb{R}^{2048\times 2}$ fed to the neural network should be normalized by appropriately shifting and scaling them:
	\begin{equation}
		\label{eq:dain-1}
		\mathbf{\tilde{x}}_j = \left(\mathbf{x}_j - \bm{\alpha} \right) \oslash \bm{\beta},
	\end{equation}
where $\mathbf{x}_j \in \mathbb{R}^2$ refers to the $j$-th observation (out of 2048 included in the current window), $\oslash$ is the Hadamard (entrywise) division operator and $\bm{\beta}$ is a scaling operator.  To this end, we first build a {summary representation} of the current window as:
\begin{equation}
	\mathbf{a} = \frac{1}{2048} \sum_{j=1}^{2048} \mathbf{x}_j \in \mathbb{R}^2.
\end{equation}
This average is used to estimate the mode of the distribution that generated the observed data. Then, DAIN learns how to appropriate shift the data based on the observed mode by estimating the value for the shifting operator $\bm{\alpha}$ as: 
\begin{equation}
	\label{eq:2}
	\bm{\alpha}=\mathbf{W}_a \mathbf{a} \in \mathbb{R}^{2},
\end{equation}
where $\mathbf{W}_a  \in  \mathbb{R}^{2 \times 2}$ are the trainable parameters of the shifting operator.

After this shifting process, the data are also scaled  by employing the scaling operator $\bm{\beta}$, as shown in (\ref{eq:dain-1}). Again, we calculate a summary representation as:
	\begin{equation}
	b\textbf{}_k = \sqrt{\frac{1}{2048} \sum_{j=1}^{2048} \left(x_{j,k} - \alpha_{k} \right)^2}, \text{\quad} k = 1, 2,
	\end{equation}
and then define the scaling operator as:
\begin{equation}
	\label{eq:3}
	\bm{\beta} = \mathbf{W}_b \mathbf{b} \in \mathbb{R}^{2},
\end{equation}
where $\mathbf{W}_b \in  \mathbb{R}^{2 \times 2}$ are its parameters.

Finally, the shifted and scaled observations are fed to an \textit{adaptive gating layer} which aims to 
appropriately 
modulate the features according to their usefulness for the task at hand as:
\begin{equation}
	\label{eq:5}
	\mathbf{\tilde{\tilde{x}}}_j = \mathbf{\tilde{x}}_j \odot \bm{\gamma},
\end{equation}
where $\odot$ is the Hadamard (entrywise) multiplication operator and
\begin{equation}
	\bm{\gamma} = \text{sigm}(\mathbf{W}_c \mathbf{c} +\mathbf{d}) \in \mathbb{R}^{2},
\end{equation}
$\text{sigm}(x) = 1 / (1 + \exp (-x) )$ is the sigmoid function,  $\mathbf{W}_c \in  \mathbb{R}^{2 \times 2}$ and $\mathbf{d} \in \mathbb{R}^{2}$ are the parameters of the gating layer that are learned through back-propagation. The updated summary representation   $\mathbf{c}$  is calculated as:
\begin{equation}
	\mathbf{c} = \frac{1}{2048} \sum_{j=1}^{2048} \mathbf{\tilde{x}}_j \in \mathbb{R}^2.
\end{equation}
The non-linearity introduced by this layer allows for the suppression of the normalized features during inference. This can help to  reduce the effect of features that could harm the generalization abilities of the network. The parameters introduced by the DAIN layer are fitted using gradient descent:
\begin{equation}
\begin{split}
		\Delta \Big( \mathbf{W}_{a}, \mathbf{W}_{b}, \mathbf{W}_{c}, \mathbf{d}, \mathbf{W} \Big) = - \eta\Big( \eta_a \frac{\partial \mathcal{L}}{\partial \mathbf{W}_{a}}, \\ \eta_b \frac{\partial \mathcal{L}}{\partial \mathbf{W}_{b}},\eta_c \frac{\partial \mathcal{L}}{\partial \mathbf{W}_{c}},\eta_c \frac{\partial \mathcal{L}}{\partial \mathbf{d}}, \frac{\partial \mathcal{L}}{\partial \mathbf{W}} \Big)
\end{split}
\end{equation}
where $\mathcal{L}$ denotes the final loss function of the network  and $\mathbf{W}$ its weights. Separate learning rates can be used for the parameters of each sub-layer, i.e. $\eta_a$,  $\eta_b$, and $\eta_c$, if necessary, to ensure the stability of the training process.

\subsection{Deep Residual Networks}

Residual neural networks \cite{he2016deep} use so-called {\it skip connections} to improve training, by allowing gradients to better reach the earliest layers of a neural network architecture, effectively solving the vanishing gradient problem \cite{hanin2018neural} and leading to more effective training as the number of layers increases, especially when {\it paired with carefully designed training methods} \cite{wightman2021resnet}. This allows for much deeper networks to be trained, in comparison to simple CNN. Depth in neural networks has been linked to higher levels of semantic information, and better performance in tasks like image recognition \cite{takahashi2017aenet}, detection \cite{xu2018deep},  etc. 

We designed a deep residual network, based on 1D convolutions, for the purpose of the binary classification of 1s long (i.e.  $2\times 2048$-dimensional)\footnote{We remind that the sampling rate was 2048Hz and there are two channels, one for each of the aLIGO detectors.} segments into positive (containing an injection) or negative (pure noise) segments. Our network has a depth of 54 layers, grouped into 27 {\it blocks} comprising two convolutional layers with a varying number of filters. The network's backbone (except for the final two convolutional layers) is summarized in Table~\ref{tab:resnet52_feature_extraction}, where $D$ denotes the dimensionality of the input tensor. Blocks 5, 8, 11, 14 and 17 are {\it 2-strided}, meaning that dimensionality is halved and an additional layer is used in the residual connection.

A graphical depiction of the network architecture is displayed in  Figure~\ref{fig:resnet52}. The output of a residual block of layers can be written  schematically as:
\begin{equation}
    \mathbf{g} = f(\mathbf{x}) + h(\mathbf{x}),
\end{equation}
where $f(\mathbf{x})$ is a block of two convolutional layers, and $\mathbf{x}$ the input (i.e. the $2\times 2048$-dimensional timeseries) to its first layer. The {\it residual function} $h(\mathbf{x})$ is either the identity function,  $h(\mathbf{x})=\mathbf{x}$, or a strided convolutional layer. Each convolutional layer is followed by batch normalization and a ReLU activation function. 
Two final individual convolutional layers gradually reduce the output to a binary outcome (noise plus injected waveform vs. noise only).


\subsection{Training}
\label{sec:training}

\vspace{0.5cm}
The network is fed with the training dataset (925k 1s segments of noise and an equal number of 1s segments containing injections) using  a mini-batch size of 400 (each mini-batch also contains an equal number of only noise and noise plus injection segments, randomly chosen from the whole dataset). The Adam optimizer \cite{https://doi.org/10.48550/arxiv.1412.6980} was used during backpropagation\footnote{Note that the first two training epochs are a warm up period, during which the learning rate is annealed from $10^{-8}$ to $4.5\times 10^{-3}$.} and the {\it regularized} binary cross entropy  (a variant of the cross entropy loss function that is designed to be finite \cite{schafer2022training})
was used as the {\it objective function}.

{\it Dynamic augmentation.} During training, we also employed dynamic augmentation,  by randomly replacing a noise segment for the L1 channel, by another segment. This augmentation is performed with a random probability of 40\% and virtually increases the total number of samples that the network sees during training. 

\begin{figure}
    \centering
    \includegraphics[width=0.48\textwidth]{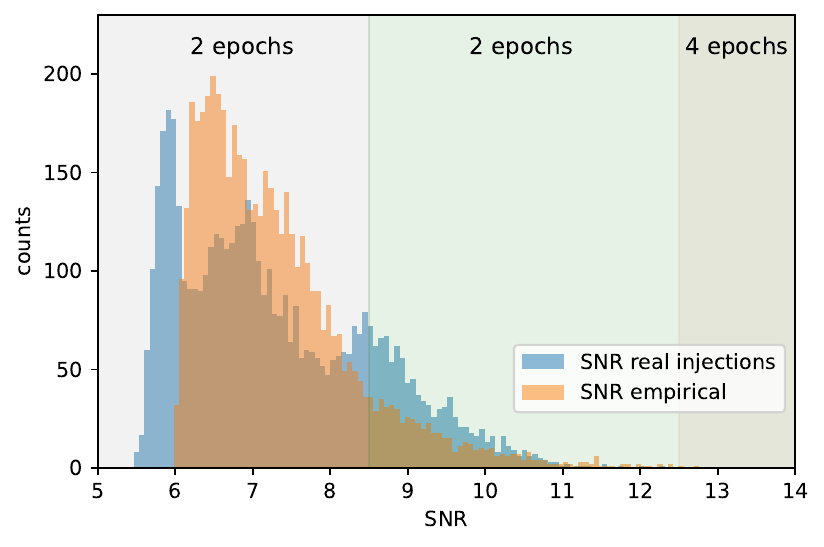}
    \caption{Comparison of the SNR histogram computed with the empirical relation Eq.~(\ref{SNR-empirical}) to the optimal SNR of the injections (using a random subset of 10000 injections). A similar distribution is obtained. The shaded areas correspond to the restricted values of the SNR used in the first eight epochs of the {\it learning strategy} (from right to left, see Table \ref{tab:training}).}
    \label{fig:SNR_comparison}
\end{figure}

\begin{figure}
    \centering
    \includegraphics[width=0.48\textwidth]{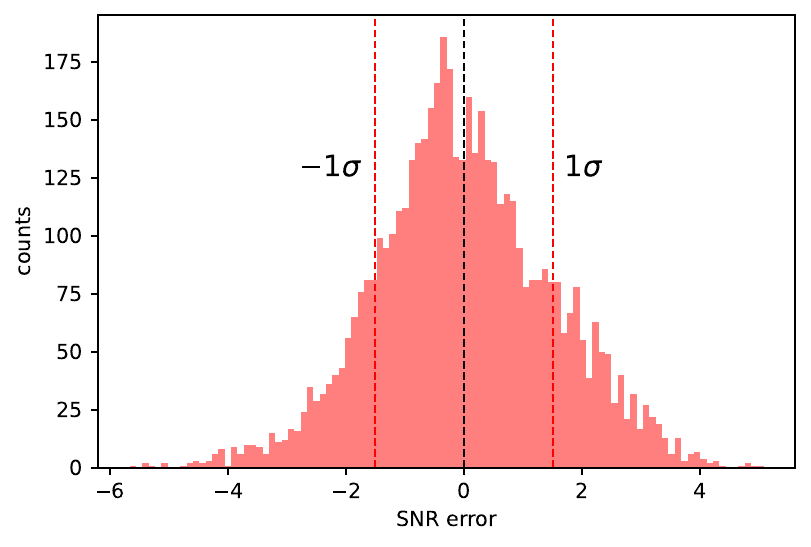}
    \caption{Histogram of the errors between the SNR computed with the empirical relation Eq. (\ref{SNR-empirical}) to the optimal SNR of the same injections as in Fig. \ref{fig:SNR_comparison}. The errors follow a normal distribution with standard deviation $\sigma = 1.1$.}
    \label{fig:norm_effect}
\end{figure}

{\it SNR-based Curriculum Learning.}  We implemented a learning strategy, so that the network is first trained on the loudest injections and only at later epochs it is trained on weaker injections. To achieve this, we used the optimal signal-to-noise ratio (SNR) of the injected signal
\begin{equation}
{\rm SNR} =2 \sqrt{ \int_0^{\infty} d f \frac{|\tilde{h}(f)|^2}{S_n(f)}},
\label{SNR-optimal}
\end{equation}
where $\tilde{h}(f)$ is the amplitude of the Fourier transform of the injected signal and $S_n(f)$ is the 
PSD of the detector's noise. An accurate computation of the SNR of each segment during training would add some computational overhead, so we chose to construct an empirical relation that gives an approximate prediction of the SNR. Specifically, by varying only the chirp mass, the distance and the inclination  $\iota$ (and setting all other parameters to some fixed values), we arrived at the following approximate empirical relation:
\begin{equation}
\mathrm{SNR}=\frac{1261 \ {\rm Mpc}}{D} \left(\frac{{\cal M}_{\rm c}}{M_\odot}\right)^{5 / 7}\left[0.7+0.3\cos (2 \ \iota)\right].
\label{SNR-empirical}
\end{equation}

Fig. \ref{fig:SNR_comparison} shows a comparison of the distribution of the optimal  SNR (obtained through Eq. (\ref{SNR-optimal}) using the PSD of the Hanford detector) for a subset of 10000 randomly chosen injections to the approximate distribution as obtained through Eq. (\ref{SNR-empirical}). The two distributions are similar, their difference stemming from the fact that the actual SNR depends on many more parameters (sky location, spins etc.). To quantify their similarity, in Fig. \ref{fig:norm_effect} we show the histogram of the errors between the two distributions (actual and empirical SNR) of  Fig. \ref{fig:SNR_comparison}. The errors follow a normal distribution with a standard deviation of $\sigma = 1.1$, which we found to be adequate for our purpose.

\begin{table}[]
    \centering
        \caption{ Learning strategy: During the first 4 epochs the network learns only strong signals with estimated optimal SNR larger than 7.63. Next, the network gradually learns weaker signals (for the subsequent epochs the estimated SNR range is shown). After the tenth epoch, the network learns all signals.}
    \label{tab:training}
    \begin{tabular}{ccc}
    \toprule
        \hspace{0.2cm} epochs \hspace{0.2cm} & \hspace{0.2cm}  min(SNR) \hspace{0.2cm} & \hspace{0.2cm} max(SNR) \hspace{0.2cm} \\ \midrule
        4 & 7.63 & 100 \\
        2 & 3.63 & 100\\
        2 & 1 & 3.63  \\
        2 & 1 & 7.63 \\
        remaining & 0 & $\infty$  \\
        \bottomrule
    \end{tabular}

\end{table}

\begin{figure}[t]
    \centering
    \includegraphics[width=0.48\textwidth]{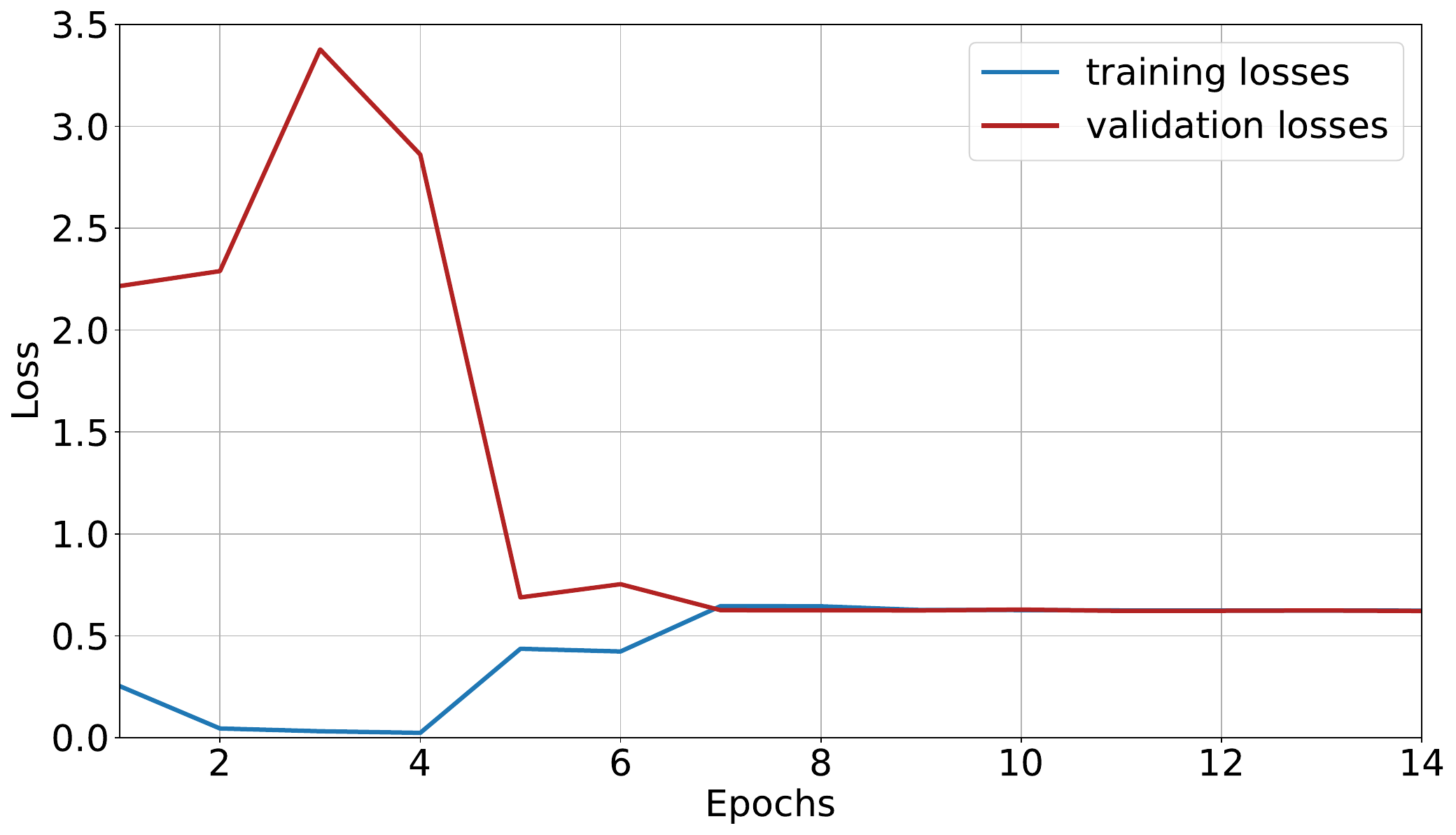}
    \caption{
     Loss as a function of epoch. Due to the adopted
learning strategy (strong signals first), initial losses are small.
After ten epochs, when the network is learning all signals, the
loss of the training set converges with that of the validation
set.
}
    \label{fig:loss}
\end{figure}

\begin{figure}[t]
    \centering
    \includegraphics[width=0.48\textwidth]{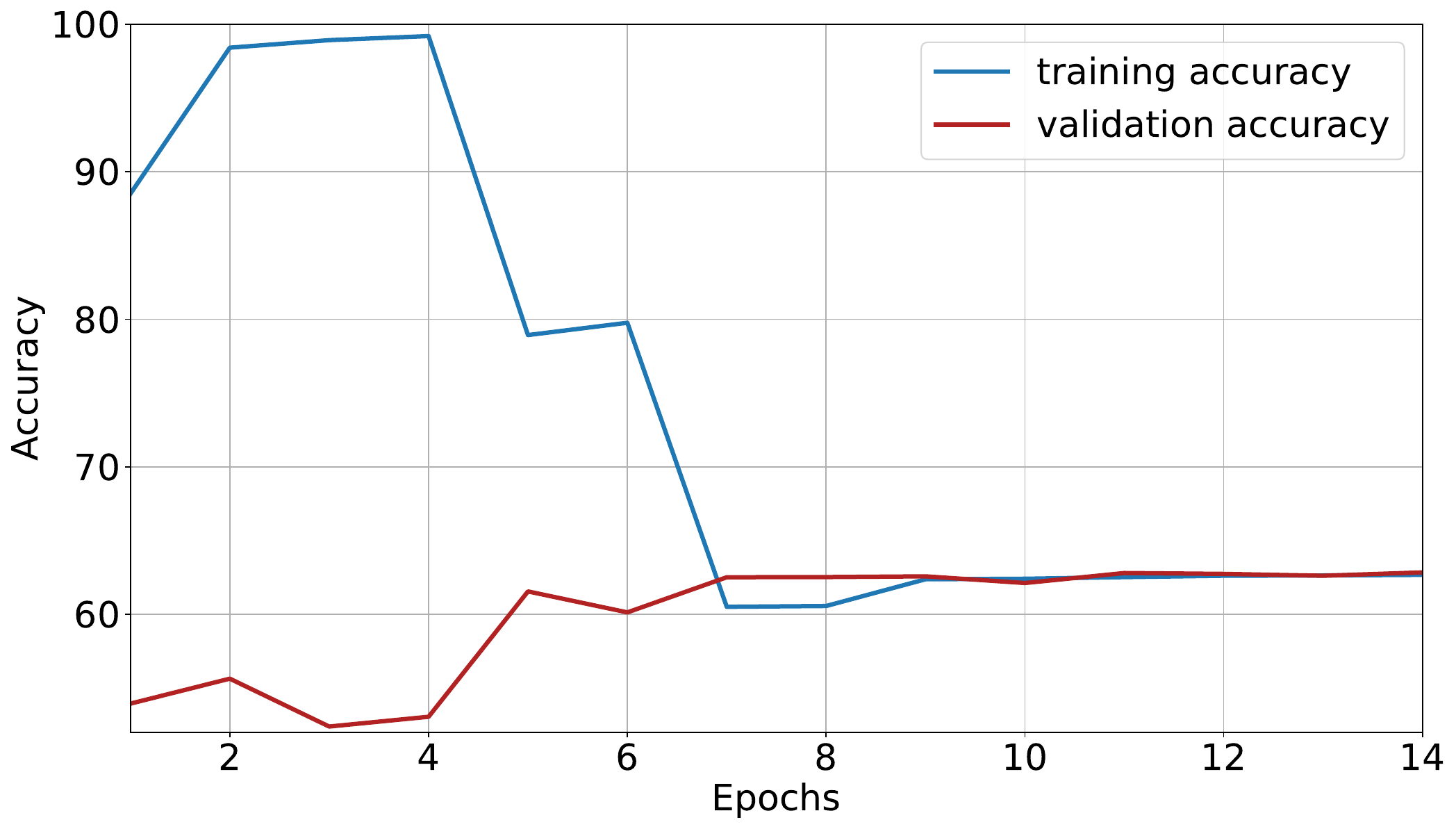}
    \caption{Accuracy as a function of epoch. Due to the adopted learning strategy (strong signals first), initial accuracy is high. After ten epochs, when the network is learning all signals, the accuracy of the training set converges with that of the validation set.}
    \label{fig:accurac}
\end{figure}

We start the training with 
easily recognizable signals with large estimated SNR, for the first four epochs. Next, the network is trained on gradually weaker signals and after the tenth period it learns all signals in the training set. The detailed SNR-based training strategy is shown in Table \ref{tab:training} and the first eight epochs are also depicted in Fig. \ref{fig:SNR_comparison} in relation to the distribution of the SNR.

Fig. \ref{fig:loss} shows the evolution of the loss function as a function of the training epoch. Due to the adopted learning strategy, initial losses are small. When the network is finally learning all signals (after ten epochs), the loss of the training set converges with that of the validation set. The corresponding evolution of the accuracy as a function of the training epoch is shown in Fig. \ref{fig:accurac}.

\begin{figure}[t]
    \centering
    \includegraphics[width=0.48\textwidth]{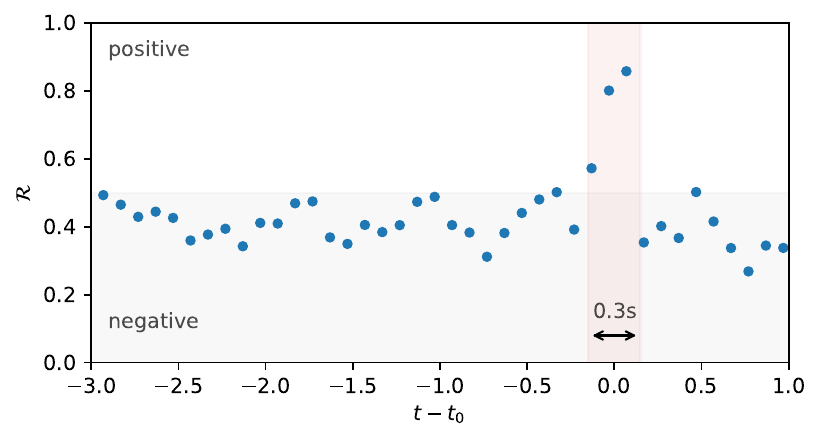}
    \caption{Representative example of the ranking statistic for overlapping segments (with a stride of 0.1s). Segments with ${\cal R}<0.5$ are classified as negatives. Segments with ${\cal R}>0.5$ that cluster within 0.3s of a known injection at time $t_0$ are reported as a single true positive.}
    \label{fig:trigger}
\end{figure}

\section{Detection of BBH injections in real noise}

The trained network is run on segments of the test dataset described in Sec. \ref{sec:datasets} and produces a binary output, corresponding to the probabilities that the waveform is present or that the segment contains only noise. The first output (waveform present) serves as a ranking statistic $\cal R$ with values between 0 and 1 for each 1s segment. When ${\cal R}>0.5$, a positive outcome is recorded. 

During deployment on the test dataset, we crop an amount of 4.25 seconds of data, from which we compute the PSD in real time. To increase speed, the Welch method for computing the PSD was implemented in PyTorch \cite{PyTorch_ref}, whereas whitening is implemented as the first layer of the final detection module. Then, the input is reorganized as a batch of 1.25s segments with a stride of 0.1s, i.e., with an overlap of 1.15s. This batch is whitened with a single forward pass of the implemented whitening module. The last valid segment starts at the 3s mark, which leads to 31 segments of duration 1.25s. 
 After whitening, the 1.25s whitened segments are cropped unilaterally to 1s total duration. Notice that when an injection is present, multiple overlapping segments can have ${\cal R}>0.5$. Because of this, we cluster positives and report any positives that are detected within a time span of 0.3s as a {\it single detection} (see Fig. \ref{fig:trigger} for a representative example).

After the deployment, the output is evaluated every 0.1s comparing it to the known injection times in the test dataset. If a positive output differs less than 0.3s from the nominal merger time for a particular injection, then it is classified as a true positive, otherwise as a false positive. 

To evaluate the effectiveness of the search algorithm, we first need to determine the false alarm rate $\cal F $ as a function of the ranking statistic, i.e. the function $\cal F (\cal R)$. Next, the sensitive distance of the search is determined as a function of the ranking statistic, which finally produces a relation between the sensitive distance and the $\cal F $ (see \cite{challenge1} for definitions and the detailed methodology).

\begin{figure} [t]
    \centering
    \includegraphics[width=0.95\linewidth]{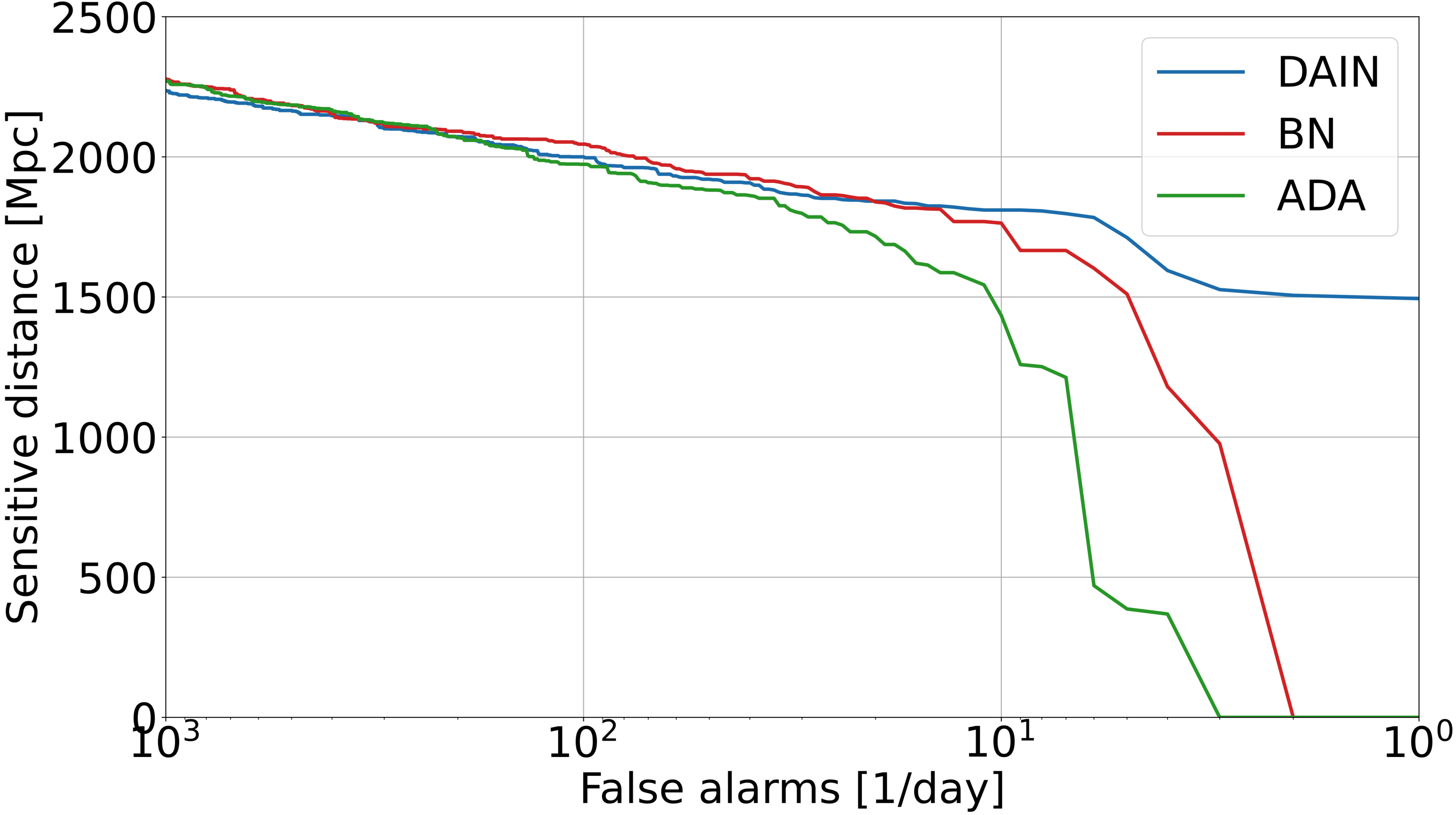}
    \caption{Sensitive distance vs. false alarm rate for three different input normalizations: DAIN \cite{dain}, Batch Normalization (BN) \cite{ioffe2015batch} and Adaptive Batch Normalization (ADA) \cite{chen2017fast}
    (for a shorter test dataset of duration of one day). In contrast to BN and ADA, DAIN maintains a good sensitive distance down to the lowest FAR.}
    \label{fig:BNADADAIN}
\end{figure}

In Fig. \ref{fig:BNADADAIN} we demonstrate the importance of using DAIN \cite{dain} as the input normalization, by comparing the sensitive distance at different FAR (for a shorter test dataset of duration of one day) to the corresponding results obtained with either Batch Normalization (BN) \cite{ioffe2015batch} or Adaptive Batch Normalization (ADA) \cite{chen2017fast}. 
Whereas the three methods achieve similar sensitive distances at FAR $> 30$/d, the performance of BN and ADA deteriorates, dropping to 0 at FAR of several per day. In contrast, the input normalization with DAIN allows the network to maintain a good sensitive distance down to the lowest FAR (notice that for the comparison shown in Fig. \ref{fig:norm_effect} we did not activate the SNR-based curriculum learning, to isolate the effect of the input normalization).

\begin{figure}[t]
    \centering
    \includegraphics[width=\linewidth]{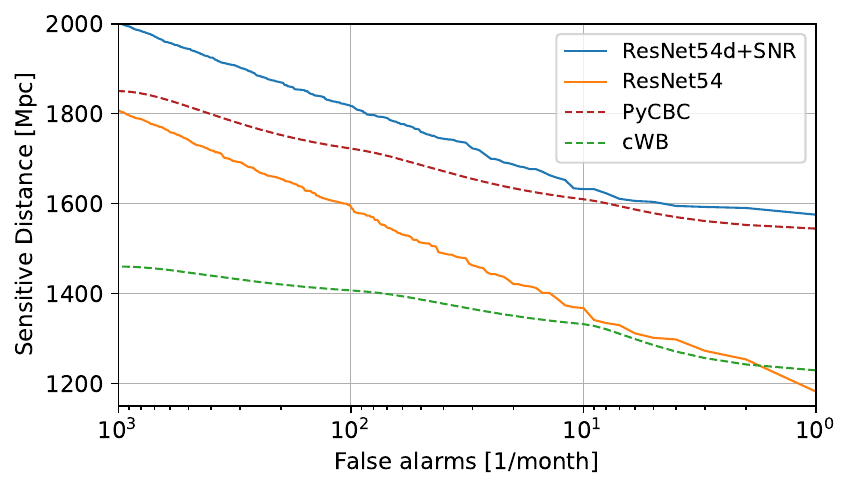}
    \caption{Sensitive distance vs. false alarm rate for our best model (ResNet54d+SNR), in comparison with the simpler setup (ResNet54) used in \cite{challenge1} and two widely used algorithms for GW detection,  Coherent WaveBurst (cWB) and PyCBC (the latter using algned-spin templates only, see text for details). All codes were run on the same test dataset established in \cite{challenge1}.  Our best model surpasses the performance of the other algorithms at all FAR in this setting (notice that the PyCBC run was based on a template bank with only aligned-spin waveforms).}
    \label{fig:comparison}
\end{figure}

Fig. \ref{fig:comparison} displays the sensitive distance as a function of FAR for our best model (ResNet54d+SNR), in comparison with a simpler setup (ResNet54\footnote{This setup has half the number of filters, as was the case in \cite{challenge1}, and does not include the SNR-based curriculum learning.}) and two widely used algorithms for GW detection,  Coherent WaveBurst (cWB), a waveform model-agnostic search pipeline for GW signals based on the constrained likelihood method \cite{PhysRevD.93.042004,cwb-softwareX,klimenko_sergey_2021_5798976} and PyCBC \cite{alex_nitz_2022_6912865}, based on a standard configuration of the archival search for compact-binary mergers \cite{2021arXiv211206878N}. The results of cWB and PyCBC shown in Fig. \ref{fig:comparison} are taken from \cite{challenge1} (where they were obtained on the same test dataset). cWB uses wavelets instead of specific models of the waveforms it searches for, which naturally does not allow it to reach ideal fitting factors. It was recently enhanced with machine-learning techniques \cite{PhysRevD.105.083018} (for details on the particular cWB setup used on the test dataset, see \cite{challenge1}). PyCBC implements matched filtering of waveform templates, but in \cite{2021arXiv211206878N,challenge1} only aligned-spin templates were used (for more general waveforms, the method could become computationally too costly). Since the injections in the test dataset are based on more general waveforms, this particular PyCBC search cannot reach ideal fitting factors, leaving room for other algorithms to surpass it.

As seen in Fig. \ref{fig:comparison}, our best model, which includes the SNR-based curriculum learning, surpasses the PyCBC results at all FAR. The sensitive distance as a function of FAR is nearly level at the lowest FAR of 1 per month, indicating that our algorithm may maintain good performance even when extended to lower FAR. Our best model also significantly exceeds the sensitivity of the unmodeled cWB search.

\section{Summary and Discussion}

We used a novel combination of ML algorithms and arrived at sensitive distances for injected BBH GW signals that surpass traditional algorithms, even at small false alarm rates. The main ingredients are a 54-layer deep residual network (ResNet), a Deep Adaptive Input Normalization (DAIN), a dynamic dataset
augmentation, and curriculum learning, based on an empirical relation for the signal-to-noise
ratio. Our best ML model surpasses the sensitive distance achieved with traditional algorithms in a specific setting that uses a dataset consisting of a large number of injected non-aligned spin waveforms in real LIGO  O3a noise samples. The matched-filtering PyCBC run on the same dataset included only aligned-spin templates and could therefore not reach optimal sensitivity.

\begin{figure}[t]
    \centering
    \includegraphics[width=\linewidth]{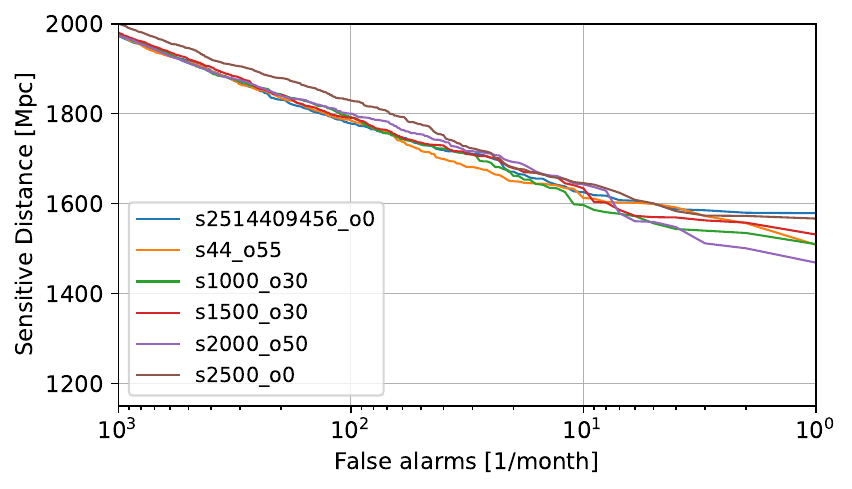}
    \caption{Variance of the sensitive distance vs. false alarm rate for our best model, using various test datasets of one month duration, that differ by the random seed used for the injections and the offset in the starting time. The first curve (seed 2514409456 and offset 0) corresponds to the case of the test dataset of the MLGWSC-1 challenge (blue curve in Fig. \ref{fig:comparison}). At a FAR of 1/month, the variance is only $\pm 3\%$ of the average.}
    \label{fig:variance_comparison}
\end{figure}

We examined the variance of our main result using additional test datasets with different offsets and starting times, see Fig. \ref{fig:variance_comparison}. The different sensitivity curves we obtain are within a few percent from the average, demonstrating that our trained neural network runs robustly on differently datasets, with consistent sensitivity.

Our ML model operates at a  fraction of the computational cost of matched filtering (see \cite{challenge1} for detailed comparisons) and it can thus be deployed in upcoming rapid online searches of gravitational-wave events in a sizeable portion of the astrophysically interesting parameter space.

The performance of our ML model could be further improved using a more accurate empirical relation for the signal-to-noise ratio and a more fine-tuned curriculum learning. We are planning to extend the training dataset to include lower or higher black hole masses. It is likely that dedicated networks will need to be trained to cover the edges of the parameter space. Furthermore, we are planning to include additional channels for the operating advanced Virgo \cite{Virgo_ref} and Kagra  \cite{KAGRA_ref} detectors. In the future, additional detectors, such as the planned LIGO-India \cite{2022CQGra..39b5004S} detector could be included in the training of the network.

Our code, {\it AResGW}, and detailed results are publicly available at \cite{AResWG}.  

\subsection{Acknowledgments}

We are grateful to Agata Trovato, Marlin Schaefer and Ondrej Zelenka for useful comments on the manuscript. We are indebted to the organizers of the MLGWSC-1 challenge, for all their work and for providing routines for creating different datasets. We acknowledge the support provided by the IT Center of the Aristotle University of Thessaloniki (AUTh) as our results have been produced using the AUTh High Performance Computing Infrastructure and Resources. We also acknowledge support from COST Action (European Cooperation in Science and Technology) CA17137 (G2Net). 

This research has made use of data or software obtained from the Gravitational Wave Open Science Center (gw-openscience.org), a service of LIGO Laboratory, the LIGO Scientific Collaboration, the Virgo Collaboration, and KAGRA. LIGO Laboratory and Advanced LIGO are funded by the United States National Science Foundation (NSF) as well as the Science and Technology Facilities Council (STFC) of the United Kingdom, the Max-Planck-Society (MPS), and the State of Niedersachsen/Germany for support of the construction of Advanced LIGO and construction and operation of the GEO600 detector. Additional support for Advanced LIGO was provided by the Australian Research Council. Virgo is funded, through the European Gravitational Observatory (EGO), by the French Centre Na- tional de Recherche Scientifique (CNRS), the Italian Istituto Nazionale di Fisica Nucleare (INFN) and the Dutch Nikhef, with contributions by institutions from Belgium, Germany, Greece, Hungary, Ireland, Japan, Monaco, Poland, Portugal, Spain. The construction and operation of KAGRA are funded by Ministry of Education, Culture, Sports, Science and Technology (MEXT), and Japan Society for the Promotion of Science (JSPS), National Research Foundation (NRF) and Ministry of Science and ICT (MSIT) in Korea, Academia Sinica (AS) and the Ministry of Science and Technology (MoST) in Taiwan.

\bibliography{AResGW_arxiv}%

\end{document}